\documentclass[a4paper,20pt]{article}
    \usepackage[T1]{fontenc}
    \usepackage{graphicx}
    \usepackage{epsfig}
    \usepackage{amsmath}
    \usepackage{amsfonts}
    \renewcommand{\abstract}{}
\begin{document}
\makeatletter
\renewcommand{\@oddhead}{\textit{Advances in Astronomy and Space Physics} \hfil \textit{A.V. Tugay}}
\renewcommand{\@evenfoot}{\hfil \thepage \hfil}
\renewcommand{\@oddfoot}{\hfil \thepage \hfil}
\fontsize{11}{11} \selectfont

\title{Extragalactic filament detection with a layer smoothing method}
\author{\textsl{A.\,V.~Tugay}}
\date{}
\maketitle
\begin{center} {\small 
Taras Shevchenko National University of Kyiv, Glushkova ave., 4, 03127, Kyiv, Ukraine\\
tugay@anatoliy@gmail.com}
\end{center}

\begin{abstract}

Filaments are clearly visible in galaxy distributions, but they are hardly detected by computer algorithms. Most methods of filament detection can be used only with numerical simulations of a large-scale structure. New simple and effective methods for the real filament detection should be developed. The method of a smoothed galaxy density field was applied in this work to SDSS data of galaxy positions. Five concentric radial layers of 100 Mpc are appropriate for filaments detection. Two methods were tested for the first layer and one more method is proposed.

\end{abstract}

\section*{Introduction}
\indent \indent Cellular large-scale structure of the Universe (LSS) can be easily seen in many galaxy distributions. It is visible in the distribution of 2MASS sources on a celestial sphere and in a number of galaxy redshift surveys. Such a structure was explained in Zeldovich theory of linear growth of fluctuations due to gravitational instability. LSS is formed under the influence of gravity from the primordial dark matter fluctuations. This process leads to the formation of such elements of LSS as domain walls, filaments, galaxy clusters and voids. All these structures were simulated on computers in many works. We can now describe LSS as a set of voids with galaxies between them. An average size of a void is 100 Mpc. The walls of the voids consist of one-dimensional filaments. Galaxy clusters, groups and isolated galaxies can be found in filaments. The largest clusters are located commonly on the intersections of the filaments, on the borders of more than two walls and voids.
 In this paper we will consider the task of filaments detection in a galaxy distribution. This task is necessary for the description of filaments and LSS as a whole. Further research in this realm can be useful for the dark matter studying and estimating of cosmological parameters. During the last years a number of methods for filament detection were developed. Most of them can be used only to numerically simulated LSS because they needs full information about the parameters of a distribution of all galaxies in a taken volume. Only three methods were applied recently to a real galaxy distribution. This set of real galaxies can be taken from Sloan Digital Sky Survey (SDSS). SDSS covers the large part of the sky of 120x70 degrees and has redshifts for up to million galaxies. Although SDSS is the best galaxy sample for filament detection, application of any computer algorithm to it is quite problematic. The main problem is lack of observed galaxies in the concrete filament. This problem leads to different nonphysical artefacts in filament detection in different methods. Nobody can recover the full network of filaments at the distances more than 500 Mpc with SDSS data. This problem was solved in \cite{sousbie11} by generating of additional galaxies between the real ones and presenting each galaxy as a complex expanded density (probability) field with its own filamentary structure. This leads to the appearance of many numerous and curved filaments and too complex and detailed shape of void bounds. Oppositely, in \cite{smith12}, large areas in the sky were not filled by filaments. The authors found only 53 filaments in all SDSS volume for redshifts z<0.15. Up to a thousand of voids and filaments should be in the such volume for the character 100 Mpc size of a void and a filament. In the paper \cite{tempel14} filaments were detected as the lines connecting nearby galaxies, groups or clusters. Real galaxy distribution has a lot of spaces between galactic structures, so this method detects a lot of small filaments instead of a single large filament.

\begin{figure}[!h]
\begin{minipage}[t]{.99\linewidth}
\centering
\epsfig{file = 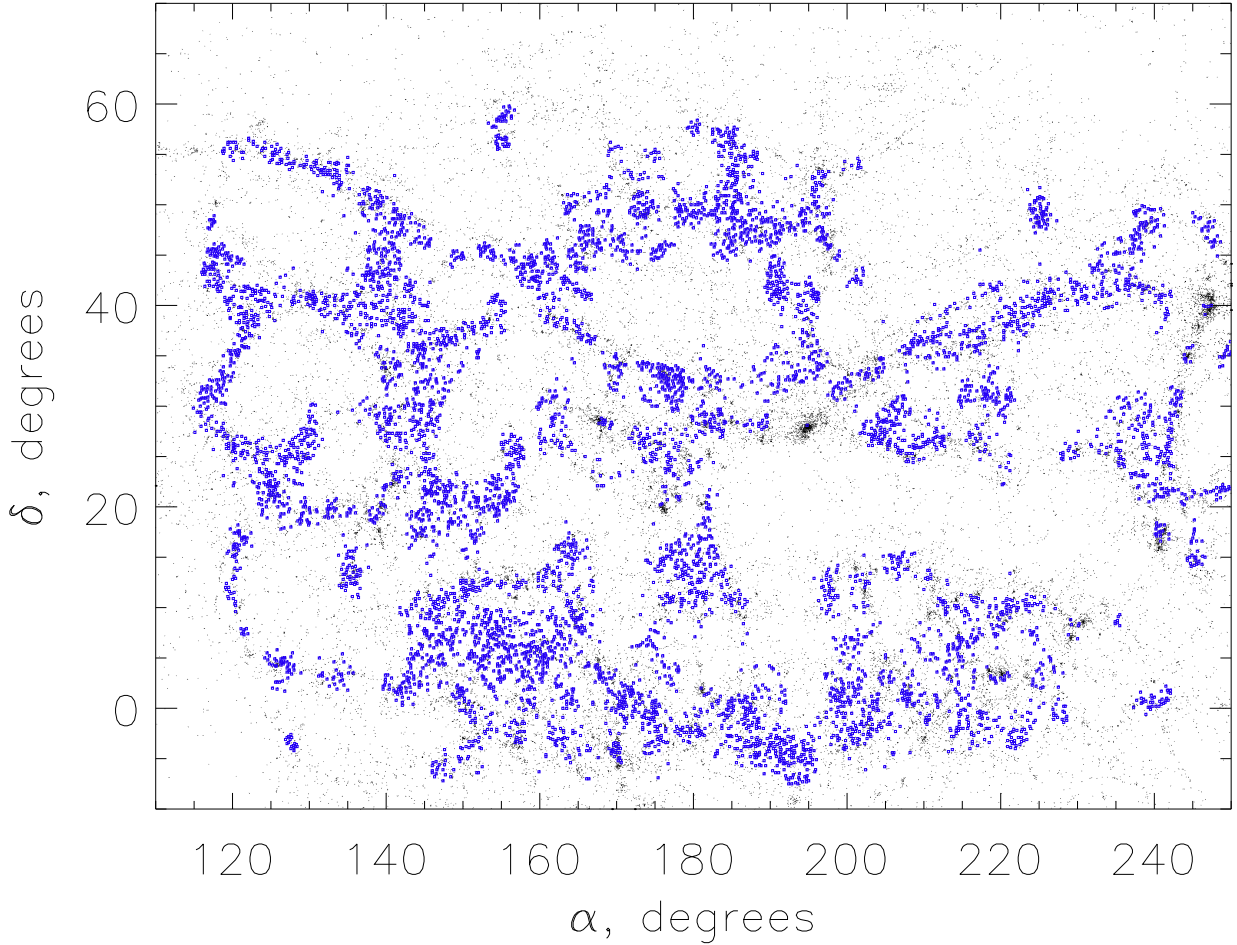, width = 0.99\linewidth}
\caption{Sky distribution of SDSS galaxies with radial velocities between 4000 and 11000 km/s. Large dots - maximums of smoothed density field.}\label{fig1}
\end{minipage}
\end{figure}

\begin{figure}[!h]
\begin{minipage}[t]{.99\linewidth}
\centering
\epsfig{file = 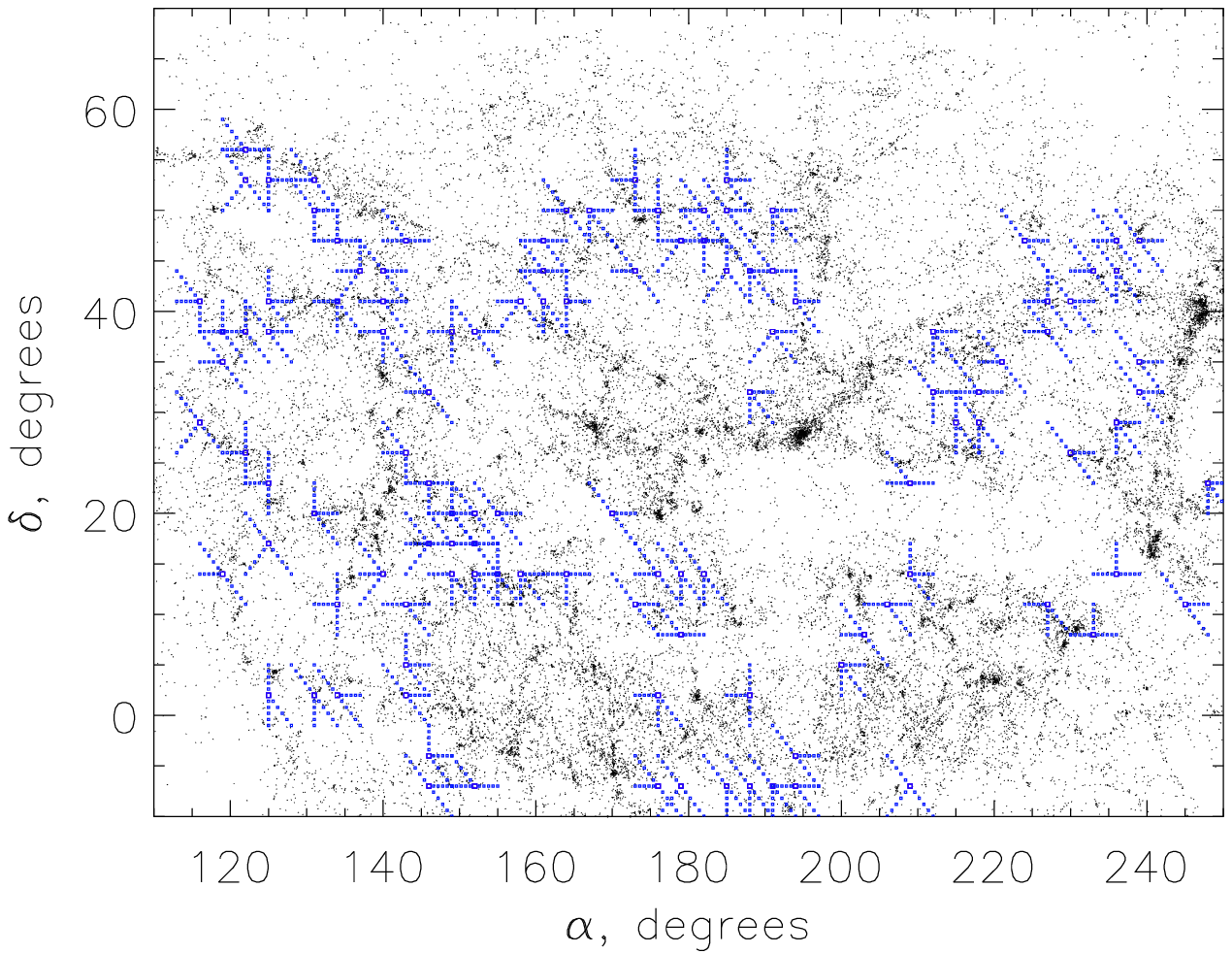, width = 0.99\linewidth}
\caption{Filament network on a square grid. Each point of a grid with smoothed density larger than a limiting value is connected with three nearby points with the closest values of galaxy density.}\label{fig2}
\end{minipage}
\end{figure}

\section*{Method}
\indent \indent The main idea is to look for filaments in concentric radial layers. The thickness of the layers was selected equal to the size of a void - 100 Mpc or 7000 km/s in radial velocity space. In 2D distribution of galaxies in such layers we can easily solve the problem of filaments intersection in the same position but at different distances in the sky. There should be no more than two filaments in the same place in a layer. Another advantage of layers consideration is a neglecting of the "finger of God" effect. Galaxies in clusters have large virial velocities, so the clusters are elongated along the line of a sight in redshift space. It is impossible to distinguish galaxies in clusters with large velocity from slow isolated galaxies which lie at different distances. Velocities of galaxies in clusters can reach up to 2000 km/s. But if we consider only the sky distribution of galaxies in a 7000 Mpc thick layer, all galaxies in a cluster will fall to the same place in one layer. A cluster can be cutted, of course, between two layers. We are going to solve this problem in the next works by moving the radial bounds of layers. 
 We will present here the results of the attempts of filament detection in the layers for a smoothed galaxy density field. The task was to fill the space between galaxies in the filament by smoothed density and leave lower density in voids. We used gaussian smoothing with two parameters: dispersion and cutoff radius. In this work galaxies are considered as uniform points. In the next papers we are going to take into account such parameters of galaxies as optical and X-ray luminosity, diameter, morphological type and spatial orientation. Four methods of further analysis of a smoothed galaxy density field are considered here.

\section*{Distribution of density maximums}
\indent \indent The problem is to find some numerical parameters of a smoothed galaxy density field, which value will indicate whether the taken direction lies in a filament or not. Also we need the parameter which will have different values in different filaments. With such parameters it will be possible to restore the network of filaments and compare it with visible distribution of galaxies in a layer. Although some filaments are visible in the layers at radial velocities up to 35000 km/s, in this paper only the first layer of SDSS galaxies is considered. Galaxies from this layer have radial velocities between 4000 and 11000 km/s. The filaments and voids of this layer are not numerous, have the largest angular sizes and are the most visually detectable. This layer includes the well known Coma cluster of galaxies. Although the filaments around Coma cluster are the most easier to study there are no definite description of them in the literature.
 At the first step the maximums of a smoothed density field were considered instead of galaxies. These maximums are less numerous than the entire galaxies and they can be sorted by the value of density. Some limiting values of density can be selected to mark maximums that traces the distribution of the filaments at the best. The distribution of maximums is shown at Fig. 1. In this work the angular resolution of 0.1 deg was used in all plots. The task of the combination of maximums into the filaments and the distinction of different filaments from one to another still remains. Note that the distribution of voids in a layer shows that the characteristic size of a void in the local Universe is less than 100 Mpc.

\section*{Minimal gradient lines}
\indent \indent Let's consider the following model of LSS presentation in the layer of smoothed density. Suppose that the intersections of filaments correspond to galaxy clusters and are the largest maximums of galaxy density. The most simple, natural and intuitively understandable case is the intersection of three filaments in each cluster. If all filaments in the layer have the same size and all angles between the intersecting filaments are equal to 120 degrees, we will have hexagonal 2D grid of filaments and voids. Such a pattern characterises the plane but not 3-dimensional space. In this toy model the lines of maximum density should pass from a cluster along the middle line of a filament to a void. Such lines pass starting from density maximums to the direction of a minimal density gradient. Also we should take less maximums than at Fig. 1. The ideal variant of the application of such a method will to take three very nearby maximums at each filament intersection. The number of maximums can be decreased by selecting the largest maximum in a circle of some fixed radius. Gradient lines that were obtained in such a way absolutely do not correspond to the distribution of visible filaments, so we will not show them here at all. The explanation of this fail may be the following. Firstly, maximums appear not only at filament intersections, but also often in any places in a filament. Secondly, even in the most optimal case, when a gradient line passes from a maximum along a filament, it always meet another line passing from the opposite end of a filament. A saddle-like shape of a smoothed density field appears in the middle of a filament and both gradient lines fall to a void. The process of the selection of the minimal gradient direction is very unstable, so almost all gradient lines fall into the voids perpendicularly to the filaments. If there were much more galaxies, the gradient lines would trace the filaments as their normal's in all points. But the real number of galaxies seems not to be enough for further development of this method. 

\section*{Tree graph on square grid}
\indent \indent The best detectable filaments are the largest ones, with sizes close to 100 Mpc. Fine filamentary structure of galaxy distribution can not be recovered because if we consider the task of detection of a small filament we will soon meet the filaments from too little number of galaxies, that are statistically unsignificant. X-ray galaxies can be observed at larger average distances than normal. Having in perspective the task of comparison of large-scale galaxy distribution in optical and X-ray band, we are interested in filament detection at larger possible distances. Thus we have to place to the basis of the method for filament detection the main features of filaments. The next method develops the idea of searching for filament intersections, that were described previously. The filament network in 2D layer can be presented as a set of intersections, some of them are connected by the straight lines (suppose the filaments are not curved). Then we have to find positions of these intersections and prepare a list of intersections which should be connected with every intersection. We should get finally a tree-like graph that covers the sky. We will consider the points of a square grid as possible intersections. The size of a grid cell should be smaller than the size of a void to avoid the skipping of real intersections. Thus the point of such a smaller grid can have any number of connections with neighbours but no more than three. The point with three connections should be the intersection of three filaments. The point with two connections will appear in the middle of the filament. The points with a single connection should be the tails of lost filaments that do not bound two intersection. Finally, the points inside the void should not have connections with neighbours at all. An example of such a network of filaments at a square grid is shown at Fig. 2. Grid lines correspond to general features of filament distribution, but the point in a filament is connected with a saw-like line instead of the straight one. 
 The plot of the entire smoothed density field leads to an unexpected feature. It was appeared that cutoff radius has much more influence on a picture than a dispersion parameter of gaussian. Moreover, the filaments are better detectable at a density map with larger dispersion but the same cutoff raduis. They are also well detectable for smoothing with a flat window function instead of gaussian. Thus in the next work we will present the distribution of galaxies as a set of clusters with same radius.

\section*{Conclusion}
\indent \indent The most perspective method for filament detection in a layer with a smoothed galaxy density field is the description of LSS as a grid of clusters with density larger than a limited value. The network of filaments can not be obtained as gradient lines of a smoothed density field.  

\section*{Acknowledgement}
\indent \indent The author is thankful to the Sloan Digital Sky Survey team. Funding for the SDSS and SDSS-II has been provided by the Alfred P.\,Sloan Foundation, the Participating Institutions, the National Science Foundation, the U.S. Department of Energy, the National Aeronautics and Space Administration, the Japanese Monbukagakusho, the Max Planck Society, and the Higher Education Funding Council for England. The SDSS Web Site is http://www.sdss.org/. The SDSS is managed by the Astrophysical Research Consortium for the Participating Institutions. The Participating Institutions are the American Museum of Natural History, Astrophysical Institute Potsdam, University of Basel, University of Cambridge, Case Western Reserve University, University of Chicago, Drexel University, Fermilab, the Institute for Advanced Study, the Japan Participation Group, Johns Hopkins University, the Joint Institute for Nuclear Astrophysics, the Kavli Institute for Particle Astrophysics and Cosmology, the Korean Scientist Group, the 
Chinese Academy of Sciences (LAMOST), Los Alamos National Laboratory, the Max-Planck-Institute for Astronomy (MPIA), the Max-Planck-Institute for Astrophysics (MPA), New Mexico State University, Ohio State University, University of Pittsburgh, University of Portsmouth, Princeton University, the United States Naval Observatory, and the University of Washington.

\end{document}